\documentclass[twocolumn,showpacs,prd]{revtex4}
\usepackage{mathrsfs}\usepackage{multirow}
\usepackage{longtable,lscape}
\usepackage{txfonts}
\usepackage{amssymb}
\usepackage{indentfirst}
\usepackage{graphicx,booktabs}
\usepackage{color}
\usepackage{amssymb}

\begin{document}
%\begin{CJK}{GBK}{}
\title{Hadronic molecules with both open charm and bottom}

\author{Zhi-Feng Sun$^{1,2}$}\email{sunzhif09@lzu.cn}
\author{Xiang Liu$^{1,2}$\footnote{Corresponding author}}\email{xiangliu@lzu.edu.cn}
\affiliation{ $^1$Research Center for Hadron and CSR Physics,
Lanzhou University and Institute of Modern Physics of CAS, Lanzhou 730000, China\\
$^2$School of Physical Science and Technology, Lanzhou University,
Lanzhou 730000,  China}
\author{Marina Nielsen}\email{mnielsen@if.usp.br}
\affiliation{Instituto de F\'{i}sica, Universidade de S\~{a}o
Paulo, C.P. 66318, 05315-970 S\~{a}o Paulo, SP, Brazil}
\author{Shi-Lin Zhu\footnote{Corresponding author}}\email{zhusl@pku.edu.cn}
\affiliation{Department of Physics and State Key Laboratory of
Nuclear Physics and Technology, Peking University, Beijing 100871,
China}

\date{\today}

\begin{abstract}
With the one-boson-exchange model, we study the interaction
between the S-wave $D^{(*)}/D^{(*)}_s$ meson and S-wave
$B^{(*)}/B^{(*)}_s$ meson considering the S-D mixing effect. Our
calculation indicates that there may exist the $B_c$-like
molecular states. We estimate their masses and list the possible
decay modes of these $B_c$-like molecular states, which may be
useful to the future experimental search.

\end{abstract}

 \maketitle
%\end{CJK}

\section{introduction}\label{sec1}

Carrying out the study of the hadron configuration beyond the
conventional $q\bar{q}$ meson and $qqq$ baryon is an intriguing
and important research topic. In the past decade, more and more
charmonium-like or bottomonium-like states were observed in the
$e^+e^-$ collision \cite{Aubert:2005rm,Aubert:2006ge,:2007sj}, $B$
meson decays
\cite{Choi:2003ue,Abe:2004zs,:2007wga,Aaltonen:2009tz} and even
$\gamma\gamma$ fusion processes
\cite{Uehara:2005qd,Uehara:2009tx,Shen:2009vs}, which have
stimulated the extensive discussion of exotic hadron
configurations (for a review see Refs.
\cite{Swanson:2006st,Zhu:2007wz,Nielsen:2009uh}).

In this work, we
report on the investigation of hadronic molecules with both open
charm and open bottom, where the interaction between the charmed
meson ($\mathcal{D}^{(*)}=[D^{(*)0},D^{(*)+},D_s^{(*)+}]$) and
bottom meson ($\mathcal{B}^{(*)}=[B^{(*)+},B^{(*)0},B_s^{(*)0}]$)
occurs via the one boson exchange (OBE). These new structures are
labeled as the $B_c$-like molecules because such systems contain
a charm ($c$) quark and an anti-bottom ($\bar{b}$) quark. Because of
the special hadron configuration, the prediction of the $B_c$-like
molecules with masses above 7 GeV can provide important
information for further experimental search at facilities such as
LHCb and the recently discussed $Z^0$ factory \cite{Chang:2010am}.

This paper is organized as follows. After the introduction, we present the formulas of effective potential of $B_c$-like molecules.
In Sec. \ref{sec3}, the numerical results are given. This work ends with the discussion and conclusion.

\section{The effective potential of $B_c$-like molecules}\label{sec2}

\begin{figure}[htb]
\centering
%\begin{tabular}{cc}
\scalebox{0.6}{\includegraphics{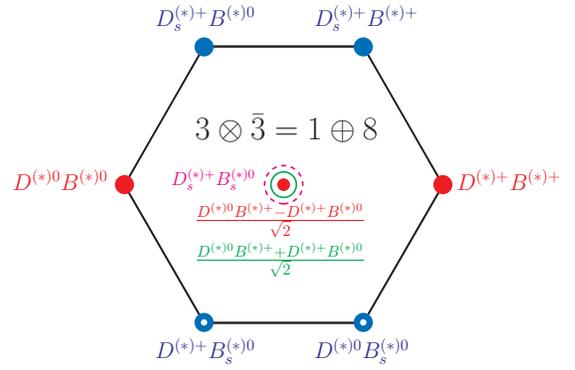}}
%\end{tabular}
\caption{(color online). The flavor wave functions of these
hadronic molecular states, which consist of two isosinglets
($a_{s1}=(D^{(*)0}B^{(*)+}+D^{(*)+}B^{(*)0})/\sqrt{2}$,
$a_{s2}=D_s^{(*)+}B_s^{(*)0}$), an isotriplet
($a_t=[D^{(*)0}B^{(*)0}$,
$(D^{(*)0}B^{(*)+}-D^{(*)+}B^{(*)0})/\sqrt{2}$,
$D^{(*)+}B^{(*)+}$]) and two isodoublets
($a_{d1}=[D_s^{(*)+}B^{(*)0},D_s^{(*)+}B^{(*)+}]$,
$a_{d2}=[D^{(*)+}B_s^{(*)0},D^{(*)0}B_s^{(*)0}]$), where the index
$a$ is taken as $X$, $Y$, $Z$ and $\widetilde{Z}$ corresponding to
the $\mathcal{DB}$, $\mathcal{D}^*\mathcal{B}^*$,
$\mathcal{D}^*\mathcal{B}$ and $\mathcal{D}\mathcal{B}^*$ systems,
respectively. \label{18}}
\end{figure}

The $B_c$-like molecules are categorized into four groups, i.e,
$\mathcal{DB}$, $\mathcal{D}^*\mathcal{B}^*$,
$\mathcal{D}^*\mathcal{B}$ and $\mathcal{D}\mathcal{B}^*$. Each
group contains nine states, which form an octet and a singlet. Their
corresponding flavor wave functions are listed in Fig. \ref{18}.
We adopt the approach developed in Refs.
\cite{Tornqvist:1993vu,Tornqvist:1993ng,Swanson:2003tb,Liu:2008fh,Liu:2008tn,Thomas:2008ja,Lee:2009hy,Sun:2011uh}
to study the interaction of the $B_c$-like molecules. In terms of
the Breit approximation, the scattering amplitude
$i\mathcal{M}({\mathcal{D}^{(*)}\mathcal{B}^{(*)}}\to
{\mathcal{D}^{(*)}\mathcal{B}^{(*)}})$ is related to the
interaction potential in the momentum space by the relation
\begin{eqnarray*}
\mathcal{V}_E(\mathbf{q})&=&\frac{-1}{\sqrt{\prod_i 2M_i \prod_f
2M_f}}\mathcal{M}\textnormal{\Huge{$[$}}
\raisebox{-13pt}{\includegraphics[width=0.075%
\textwidth]{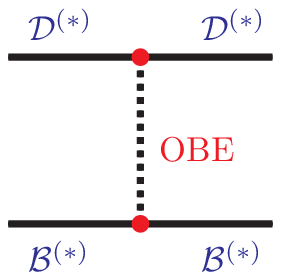}}\textnormal{\Huge{$]$}},
\end{eqnarray*}
where $M_{i}$ and $M_j$ are the masses of the initial and final
states, respectively. The potential in the coordinate space
$\mathcal{V}(\mathbf{r})$ reads as its Fourier transformation,
\begin{eqnarray}
\mathcal{V}_E(\mathbf{r})=\int\frac{d\mathbf{p}}{(2\pi)^3}\,e^{i
\mathbf{p}\cdot
\mathbf{r}}\;\mathcal{V}_E(\mathbf{q})\;\mathcal{F}^2(q^2,m_E^2)
\end{eqnarray}
where $m_E$ is the exchanged meson mass and the monopole form
factor (FF)
$\mathcal{F}(q^2,m_E^2)=({\Lambda^2-m_E^2})/({\Lambda^2-q^2})$ is
introduced to depict the structure effect of the vertex of the
heavy mesons interacting with the light mesons. The parameter
$\Lambda$, which is about one to several GeV, not only denotes the
phenomenological cutoff, but also regulates the effective
potential.

According to the heavy quark limit and chiral symmetry, the
interactions of the light pesudoscalar, vector and scalar mesons
with the S-wave heavy flavor mesons were constructed as
\cite{Cheng:1992xi,Yan:1992gz,Wise:1992hn,Burdman:1992gh,Casalbuoni:1996pg,Falk:1992cx}
\begin{eqnarray}
\mathcal{L}_{HH\mathbb{P}}&=& ig\langle H^{(Q)}_b \gamma_\mu
A_{ba}^\mu\gamma_5 \bar{H}^{(Q)}_a\rangle \nonumber\\&&+ ig\langle
\bar{H}^{(\bar{Q})}_a \gamma_\mu A_{ab}^\mu\gamma_5H_b^{(\bar{Q})}\rangle,\label{eq:lag}\\
\mathcal{L}_{HH\mathbb{V}}&=& i\beta\langle H^{(Q)}_b v_\mu
(\mathcal{V}^\mu_{ba}-\rho^\mu_{ba})\bar{H}^{(Q)}_a\rangle\nonumber\\&&+i\lambda\langle
H^{(Q)}_b
\sigma_{\mu\nu}F^{\mu\nu}(\rho)\bar{H}^{(Q)}_a\rangle\nonumber\\
&&-i\beta\langle \bar{H}^{(\bar{Q})}_a v_\mu
(\mathcal{V}^\mu_{ab}-\rho^\mu_{ab})H^{(\bar{Q})}_b\rangle\nonumber\\&&+i\lambda\langle
H_b^{(\bar{Q})}
\sigma_{\mu\nu}F'^{\mu\nu}(\rho)\bar{H}^{(\bar{Q})}_a\rangle,\\
\mathcal{L}_{ HH\sigma}&=&g_s \langle H^{(Q)}_a\sigma
\bar{H}^{(Q)}_a\rangle+g_s \langle \bar{H}^{(\bar{Q})}_a\sigma
H^{(\bar{Q})}_a\rangle,\label{eq:lag2}
\end{eqnarray}
where the multiplet fields are expressed as
$H_a^{(Q)}=\frac{1+\rlap\slash
v}{2}[{\mathcal{P}}^{*}_{a\mu}\gamma^\mu
    -{\mathcal{P}}_a\gamma_5]$, $H^{(\bar Q)}=[\widetilde{P}_a^{*\mu}\gamma_\mu-\widetilde{P}_a\gamma_5]\
    \frac{1-\rlap\slash v}{2}$, $\bar{H}=\gamma_0H^\dag\gamma_0$ with $v=(1,{\mathbf 0})$,
${\mathcal{P}}^{(*)T} =(D^{(*)0},D^{(*)+},D_s^{(*)+})$ or
$(B^{(*)-},\bar{B}^{(*)0},\bar{B}_s^{(*)0})$,
$\widetilde{{\mathcal{P}}}^{(*)T}
=(\bar{D}^{(*)0},D^{(*)-},D_s^{(*)-})$ or
$(B^{(*)+},{B}^{(*)0},{B}_s^{(*)0})$, which satisfy the
normalization relations $\langle
0|{\mathcal{P}}|Q\bar{q}(0^-)\rangle =\langle
0|\widetilde{\mathcal{P}}|\bar{Q}q(0^-)\rangle
=\sqrt{M_\mathcal{P}}$ and $\langle
0|{\mathcal{P}}^*_\mu|Q\bar{q}(1^-)\rangle=\langle
0|\widetilde{\mathcal{P}}^{*}_\mu|\bar{Q}q(1^-)\rangle=\epsilon_\mu\sqrt{M_{\mathcal{P}^*}}$.
The axial current reads as
$A^\mu=\frac{1}{2}(\xi^\dag\partial_\mu\xi-\xi \partial_\mu
\xi^\dag)=\frac{i}{f_\pi}\partial_\mu{\mathbb P}+\cdots$ with
$\xi=\exp(i\mathbb{P}/f_\pi)$ and $f_\pi=132$ MeV.
$\rho^\mu_{ba}=ig_{V}\mathbb{V}^\mu_{ba}/\sqrt{2}$,
$F_{\mu\nu}(\rho)=\partial_\mu\rho_\nu - \partial_\nu\rho_\mu +
[\rho_\mu,{\ } \rho_\nu]$, $F'_{\mu\nu}(\rho)=\partial_\mu\rho_\nu
-
\partial_\nu\rho_\mu -
[\rho_\mu,{\ } \rho_\nu]$ and $g_V=m_\rho/f_\pi$, with $g_V=5.8$.
In the above expressions, $\mathbb P$ and $\mathbb V$ denote the
three by three pseudoscalar and
vector matrices, respectively, {\it i.e.},
\begin{eqnarray}
\mathbb{P}&=&\left(\begin{array}{ccc}
\frac{\pi^{0}}{\sqrt{2}}+\frac{\eta}{\sqrt{6}}&\pi^{+}&K^{+}\\
\pi^{-}&-\frac{\pi^{0}}{\sqrt{2}}+\frac{\eta}{\sqrt{6}}&
K^{0}\\
K^- &\bar{K}^{0}&-\frac{2\eta}{\sqrt{6}}
\end{array}\right),\\
\mathbb{V}&=&\left(\begin{array}{ccc}
\frac{\rho^{0}}{\sqrt{2}}+\frac{\omega}{\sqrt{2}}&\rho^{+}&K^{*+}\\
\rho^{-}&-\frac{\rho^{0}}{\sqrt{2}}+\frac{\omega}{\sqrt{2}}&
K^{*0}\\
K^{*-} &\bar{K}^{*0}&\phi
\end{array}\right).\label{vector}
\end{eqnarray}
% $\langle\cdots\rangle$ .
The coupling constants involved in Eqs.
(\ref{eq:lag})-(\ref{eq:lag2}) include $g=0.59$ extracted from the
experimental width of $D^{*+}$ ~\cite{Isola:2003fh}, $\beta=0.9$
determined by the vector meson dominance mechanism, $\lambda=0.56$
GeV$^{-1}$ obtained by comparing the form factor calculated by
light cone sum rule with the one obtained by lattice QCD. In
addition, the coupling constant related to the scalar meson
$\sigma$, $g_s=g_\pi/(2\sqrt{6})$ with $g_\pi=3.73$ was given in
Ref. \cite{Falk:1992cx}. In the heavy quark limit, the
interactions of the $\mathcal{D}^{(*)}\mathcal{D}^{(*)}$ and
$\mathcal{B}^{(*)}\mathcal{B}^{(*)}$ with light mesons are the
same.

With these Lagrangians listed in Eq.
(\ref{eq:lag})-(\ref{eq:lag2}), we can deduce the expressions of
$\mathcal{V}_E(\mathbf{r})$. When obtaining the total effective
potentials, we sandwich $\mathcal{V}_E(\mathbf{r})$ between the
corresponding $B_c$-like molecular states. Thus, the general
expression of the total effective potential is expressed as
\begin{eqnarray}
\mathcal{V}_{Total}^{\,\,a_\xi}(\mathbf{r})=\Big\langle
a_\xi[J]\Big|\sum_{E=\pi,\eta,\rho,\omega,...}\mathcal{V}_E(\mathbf{r})\Big|a_\xi[J]\Big\rangle,
\end{eqnarray}
where subscript $a_\xi$ with $\xi=s1,s2,t,d1,d2$ and
$a=X,Y,Z,\widetilde{Z}$ is introduced to distinguish the total
effective potentials of the molecular systems defined in Fig.
\ref{18}. $J$ denotes the total angular momentum of system ($J=0$,
$J=1$, $J=0,1,2$ for the $\mathcal{DB}$,
$\mathcal{D^*B}/\mathcal{DB}^*$ and $\mathcal{D^*B}^*$ systems
respectively). The definitions of $\left|a_\xi[J]\right\rangle$
are
\begin{eqnarray*}
\left|X_\xi[0]\right\rangle&=&\left|\mathcal{DB}(^1S_0)\right\rangle,\\
\left|Z_\xi[1]\right\rangle&=&\left(\left|\mathcal{D^*B}(^3S_1)\right\rangle,\left|\mathcal{D^*B}(^3D_1)\right\rangle\right)^T,\\
\left|\widetilde{Z}_\xi[1]\right\rangle&=&\left(\left|\mathcal{DB^*}(^3S_1)\right\rangle,\left|\mathcal{DB^*}(^3D_1)\right\rangle\right)^T,\\
\left|Y_\xi[0]\right\rangle&=&\left(\left|\mathcal{D^*B^*}(^3S_0)\right\rangle,\left|\mathcal{D^*B^*}(^5D_0)\right\rangle\right)^T,\\
\left|Y_\xi[1]\right\rangle&=&\left(\left|\mathcal{D^*B^*}(^3S_1)\right\rangle,\left|\mathcal{D^*B^*}(^3D_1)\right\rangle,\left|\mathcal{D^*B^*}(^5D_1)\right\rangle\right)^T,\\
\left|Y_\xi[2]\right\rangle&=&\big(\left|\mathcal{D^*B^*}(^5S_2)\right\rangle,\left|\mathcal{D^*B^*}(^1D_2)\right\rangle,\left|\mathcal{D^*B^*}(^3D_2)\right\rangle,\nonumber\\&&\left|\mathcal{D^*B^*}(^5D_2)\right\rangle\big)^T,
\end{eqnarray*}
with
\begin{eqnarray}
    &&\Big|\mathcal{DB}^*/\mathcal{D}^*\mathcal{B}\big(^{2S+1}L_J\big)\Big\rangle=\sum_{m,m_L,m_S}C_{1m,L m_L}^{JM}
    |\epsilon^{m}_{n}; L,m_L\rangle,\label{po0}\\
   &&\Big|\mathcal{D}^*\mathcal{B}^*\big(^{2S+1}L_J\big)\Big\rangle=\sum_{m,m',m_L,m_S}C_{Sm_S,Lm_L}^{JM}
    C_{1m,1m'}^{Sm_S}|\epsilon^{m'}_{n'}\epsilon^{m}_{n}; L, m_L\rangle,\label{po}\nonumber\\
\end{eqnarray}
where $C_{1m,Lm_L}^{JM}$, $C_{Sm_S,Lm_L}^{JM}$ and
$C_{1m,1m'}^{Sm_S}$ denote the Clebsch-Gordan coefficients.
%$Y_{Lm_L}$ is the spherical harmonics function.
%\textcolor[rgb]{0.00,0.50,1.00}{As an example, we calculate the matrix element of the operator $\mbox{\boldmath$\epsilon$}\cdot \mbox{\boldmath$\epsilon$}^{\prime \dag}$}
%\begin{eqnarray*}
%\langle \mbox{\boldmath$\epsilon$}\cdot \mbox{\boldmath$\epsilon$}^{\prime \dag} \rangle &=& \sum_{\lambda^\prime,\lambda}\langle %1,\lambda^\prime;L^\prime,-\lambda^\prime|1,0\rangle \langle 1,\lambda;L,-\lambda|1,0\rangle\\
%&&\times \langle Y_{L^\prime,-\lambda^\prime}|\mbox{\boldmath$\epsilon$}^\lambda\cdot \mbox{\boldmath$\epsilon$}^{\prime \lambda^\prime %\dag}|Y_{L,-\lambda}\rangle\\
%&=&\sum_{\lambda^\prime,\lambda}\langle 1,\lambda^\prime;L^\prime,-\lambda^\prime|1,0\rangle \langle 1,\lambda;L,-\lambda|1,0\rangle \\
%&&\langle Y_{L^\prime,-\lambda^\prime}|Y_{L,-\lambda}\rangle \delta_{\lambda^\prime,\lambda}\\
%&=&\sum_{\lambda} \langle 1,\lambda;L^\prime,-\lambda|1,0 \rangle \langle 1,\lambda; L, -\lambda|1,0 \rangle \delta_{L,L^\prime}.
%\end{eqnarray*}
The polarization
vector for the vector heavy flavor meson is written as
$\epsilon^m_\pm=\mp\frac{1}{\sqrt{2}}(\epsilon^m_x\pm
i\epsilon^m_y)$ and $\epsilon^m_0=\epsilon^m_z$. In the above
expressions, $^{2S+1}L_J$ is applied to denote the total spin $S$,
angular momentum $L$, total angular momentum $J$ of the
$\mathcal{D}^{(*)}\mathcal{B}^{(*)}$ systems, while
$L=S$ and $L=D$ are introduced to distinguish S-wave and D-wave
interactions. Because of the S-D mixing effect, the obtained total
effective potentials of the $\mathcal{D^*B}$, $\mathcal{DB^*}$ and
$\mathcal{D^*B^*}$ molecular systems are in matrix form. The total
effective potentials are composed of subpotentials as shown in
Table. \ref{potential}.

\renewcommand{\arraystretch}{1.6}
\begin{table}[htb]
    \caption{The relation of the total effective potential $\mathcal{V}_{Total}^{\,\,a_\xi}(r)$
    and the subpotentials. Here, $\varpi$ is taken as 3 and -1 corresponding to the states
    marked by the subscripts $s1$ and $t$, respectively. Since the total effective potential
    of the $\mathcal{DB}^*$ systems is the same as that of the $\mathcal{D^*B}$ systems, we only show the
    result for $\mathcal{D^*B}$. We use $-$ to denote the case when the OBE potential does not exist since
    no suitable meson exchange is allowed for these systems.
    \label{potential}}
\small
\renewcommand\tabcolsep{0.08cm}
\begin{center}
    \begin{tabular}{c|cc cccccc}  \toprule[1pt]
%\multicolumn{6}{c} {{\it Established states}}    \\
$a_\xi$&$X_{s1}/X_{t}$&$X_{s2}$&$X_{d1}/X_{d2}$
\\\midrule[0.5pt]
$\mathcal{V}_{Total}^{\,\,a_\xi}(r)$&$V_{\sigma}^{\mathcal{DB}}+\frac{\varpi}{2}V_{\rho}^{\mathcal{DB}}+\frac{1}{2}V_{\omega}^{\mathcal{DB}}$
&$V_{\phi}^{\mathcal{DB}}$& $-$
\\\midrule[1pt]\midrule[1pt]
$a_\xi$&$Y_{s1}/Y_{t}$&$Y_{s2}$&$Y_{d1}/Y_{d2}$
\\\midrule[0.5pt]
\multirow{2}*{$\mathcal{V}_{Total}^{\,\,a_\xi}(r)$}&$V_{\sigma}^{\mathcal{D^*B^*}}+\frac{\varpi}{2}V_{\rho}^{\mathcal{D^*B^*}}
+\frac{1}{2}V_{\omega}^{\mathcal{D^*B^*}}$
&\multirow{2}*{$V_{\phi}^{\mathcal{D^*B^*}}+\frac{2}{3}V_{\eta}^{\mathcal{D^*B^*}}$}&
\multirow{2}*{$-\frac{2}{3}V_{\eta}^{\mathcal{D^*B^*}}$}\\
&$+\frac{\varpi}{2}V_{\pi}^{\mathcal{D^*B^*}}+\frac{1}{6}V_{\eta}^{\mathcal{D^*B^*}}$
&&\\\midrule[1pt]\midrule[1pt] $a_\xi$&$Z_{s1}/Z_{t}$&$Z_{s2}$&$Z_{d1}/Z_{d2}$
\\\midrule[0.5pt]
$\mathcal{V}_{Total}^{\,\,a_\xi}(r)$&$V_{\sigma}^{\mathcal{D^*B}}+\frac{\varpi}{2}V_{\rho}^{\mathcal{D^*B}}+\frac{1}{2}V_{\omega}^{\mathcal{D^*B}}$
&$V_{\phi}^{\mathcal{D^*B}}$& $-$
\\
\bottomrule[1pt]
\end{tabular}
\end{center}
\end{table}

The expressions of the subpotentials are
\begin{eqnarray}
V^{\mathcal{DB}}_{\sigma}&=&-g_s^2Y(\Lambda,m_\sigma,r),\quad V^{\mathcal{DB}}_{_{\mathbb{V}}}=-\frac{1}{2}\beta^2g_V^2Y(\Lambda,m_{_{\mathbb{V}}},r),\nonumber\\
V^{\mathcal{D^*B}}_{\sigma}&=&-g_s^2Y(\Lambda,m_\sigma,r)\,\mathrm{diag}(1,1),\nonumber\\
V^{\mathcal{D^*B}}_{_{\mathbb{V}}}&=&-\frac{1}{2}\beta^2g_V^2 Y(\Lambda,m_{_{\mathbb{V}}},r)\,\mathrm{diag}(1,1),\nonumber\\
V^{\mathcal{D^*B^*}}_{\sigma}&=&-g_s^2\mathcal{A}[J]Y(\Lambda,m_\sigma,r),\nonumber\\
V^{\mathcal{D^*B^*}}_{_{\mathbb{V}}}&=&-\frac{1}{4}\bigg\{2\beta^2g_V^2\mathcal{A}[J]-8\lambda^2g_V^2\bigg(\frac{2}{3}\mathcal{B}[J] \nabla^2\nonumber\\&&-\frac{1}{3}\mathcal{C}[J]r\frac{\partial}{\partial r}\frac{1}{r}\frac{\partial}{\partial r}\bigg)\bigg\}Y(\Lambda,m_{_{\mathbb{V}}},r),\nonumber\\
V^{\mathcal{D^*B^*}}_{_{\mathbb{P}}}&=&-\frac{g^2}{f_\pi^2}\bigg[\frac{1}{3}\mathcal{B}[J]\nabla^2+\frac{1}{3}\mathcal{C}[J]
r\frac{\partial}{\partial r}\frac{1}{r}\frac{\partial}{\partial
r}\bigg]Y(\Lambda,m_{_{\mathbb{P}}},r)\nonumber
\end{eqnarray}
with $Y(\Lambda,m_E,r) = \frac{1}{4\pi r}(e^{-m_E\,r}-e^{-\Lambda
r})-\frac{\Lambda^2-m_E^2}{8\pi \Lambda }e^{-\Lambda r},$ where we
use superscripts $\mathcal{D}^{(*)}\mathcal{B}^{(*)}$ to
distinguish these subpotentials for the different systems while
the introduced subscripts $\mathbb{P}$ and $\mathbb{V}$ denote the
corresponding light pseudoscalar and vector meson exchanges
respectively. $m_E$ denotes the mass of exchange meson. Matrices
$\mathcal{A}[J]$, $\mathcal{B}[J]$ and $\mathcal{C}[J]$ are listed
below with $\mathcal{A}[0]=\mathrm{diag}(1,1)$,
$\mathcal{A}[1]=\mathrm{diag}(1,1,1)$,
$\mathcal{A}[2]=\mathrm{diag}(1,1,1,1)$,
$\mathcal{B}[0]=\mathrm{diag}(2,-1)$,
$\mathcal{B}[1]=\mathrm{diag}(1,1,-1)$,
$\mathcal{B}[2]=\mathrm{diag}(-1,1,1,-1)$,
$\mathcal{C}[0]=\left(\begin{array}{cc}
0&\sqrt{2}\\
\sqrt{2}&2\\
\end{array}\right)$, $\mathcal{C}[1]=\left(\begin{array}{ccc}
0&-\sqrt{2}&0\\
-\sqrt{2}&1&0\\
0&0&1\\
\end{array}\right)$ and $\mathcal{C}[2]=\left(\begin{array}{cccc}
0&\sqrt{\frac{2}{5}}&0&-\sqrt{\frac{14}{5}}\\
\sqrt{\frac{2}{5}}&0&0&-\frac{2}{\sqrt{7}}\\
0&0&-1&0\\
-\sqrt{\frac{14}{5}}&-\frac{2}{\sqrt{7}}&0&-\frac{3}{7}\\
\end{array}\right)$. In addition, the kinetic terms for the $\mathcal{B^{(*)}D^{(*)}}$ systems are
\begin{eqnarray*}
    K_{\mathcal{DB}}&=&-\frac{\triangle}{2\tilde{m}_1},\\
    K_{\mathcal{D^*B}/\mathcal{DB^*}}&=&\mathrm{diag}\Bigg(-\frac{\triangle}{2\tilde{m}_2},~
    -\frac{\triangle_2}{2\tilde{m}_2}\Bigg),\\
    K_{\mathcal{B^*D^*}[J=0]}&=&\mathrm{diag}\Bigg(-\frac{\triangle}{2\tilde{m}_3},~
    -\frac{\triangle_2}{2\tilde{m}_3}\Bigg),\\
    K_{\mathcal{B^*D^*}[J=1]}&=&\mathrm{diag}\Bigg(-\frac{\triangle}{2\tilde{m}_3},~
    -\frac{\triangle_2}{2\tilde{m}_3},~
    -\frac{\triangle_2}{2\tilde{m}_3}\Bigg),\\
K_{\mathcal{B^*D^*}[J=2]}&=&\mathrm{diag}\Bigg(-\frac{\triangle}{2\tilde{m}_3},~
    -\frac{\triangle_2}{2\tilde{m}_3},~
    -\frac{\triangle_2}{2\tilde{m}_3},~
    -\frac{\triangle_2}{2\tilde{m}_3}\Bigg),
\end{eqnarray*}
where $\triangle=\frac{1}{r^2}\frac{\partial}{\partial
r}r^2~\frac{\partial}{\partial r}$, $\triangle_2=\triangle
-{6\over{r^2}}$. $\tilde{m}_1$, $\tilde{m}_2$ and $\tilde{m}_3$
are the reduced masses of the corresponding systems.

\section{numerical result}\label{sec3}

With the above preparation, in the following we illustrate the
numerical results for the $B_c$-like molecular systems. In order
to obtain the information of the bound-state solutions (binding
energy and root-mean-square radius) of systems listed in
Fig. \ref{18}, we need to solve the coupled channel
Schr\"{o}dinger equation with the deduced effective potentials,
which can answer whether these $B_c$-like molecular states exist
or not. Here, we adopt FESSDE, a Fortran program for solving the
coupled channel Schr\"{o}dinger equation
\cite{Abrashkevich1995,Abrashkevich1998}, to numerically obtain
the binding energy and the corresponding root-mean-square 
radius. Additionally we also use a MATLAB package MATSCE
\cite{matscs} to do a cross-check. Usually the OBE potential is
suitable to describe the interaction of a loosely bound state.
Thus, we require the obtained binding energy in the range of
$0\sim -20$ MeV and the cutoff in the range of $1\sim 5$ GeV when
presenting the result.

\begin{table*}[htbp]
\begin{center}
%\vspace{-3mm}
%\footnotesize
\caption{The typical values of the obtained bound-state solutions
for the $\mathcal{D^{(*)}B^{(*)}}$ systems. Here, $\Lambda$, $E$,
and $r_{\mathrm{RMS}}$ are in units of GeV, MeV, and fm,
respectively. \label{tp}}
\begin{tabular}{ccccc|cccccccccccccccccc}
\toprule[1pt] System&State& $\Lambda/E/r_{\mathrm{RMS}}$&State&
$\Lambda/E/r_{\mathrm{RMS}}$&System&State&
$\Lambda/E/r_{\mathrm{RMS}}$
\\\midrule[0.5pt]
%\multicolumn{8}{c}{\textbf{$\mathcal{DB}$}} \\\midrule[1pt]
\multirow{3}*{$\mathcal{DB}$}&\multirow{3}*{$X_{s1}$}&1.3/-1.28/2.58& \multirow{3}*{$X_{s2}$} &3.2/-2.03/1.99&\multirow{6}*{$\mathcal{D^*B}$}& \multirow{3}*{$Z_{s1}$}&1.3/-2.10/2.07\\
                  &&1.4/-6.14/1.35&                              &4.0/-10.87/0.94&&&1.4/-7.92/1.19                         \\
                  &&1.5/-13.91/0.98&                             &4.8/-21.90/0.69&&&1.5/-16.69/0.89\\
\cline{1-5}
\multirow{3}*{$\mathcal{DB^*}$}     &\multirow{3}*{$\tilde{Z}_{s1}$}&1.3/-1.32/2.60&\multirow{3}*{$\tilde{Z}_{s2}$}&3.0/-0.76/3.12&& \multirow{3}*{$Z_{s2}$} &3.0/-1.83/2.04\\
                          &&1.4/-6.21/1.34&                                &3.5/-4.96/1.32&&&3.5/-7.55/1.08\\
                          &&1.5/-14.02/0.97&                               &4.0/-11.06/0.93&  &                      &4.0/-15.08/0.80\\
\midrule[0.5pt]
\end{tabular}
\begin{tabular}{c|cccccccccccccccccccccc}
\toprule[1pt]
System&State& $\Lambda/E/r_{\mathrm{RMS}}$&State& $\Lambda/E/r_{\mathrm{RMS}}$&State& $\Lambda/E/r_{\mathrm{RMS}}$&State& $\Lambda/E/r_{\mathrm{RMS}}$&State& $\Lambda/E/r_{\mathrm{RMS}}$\\
\midrule[0.5pt]
\multirow{9}*{$\mathcal{D^*B^*}$}&\multirow{3}*{$Y_{s1}^{J=0}$}&1.25/-2.70/2.00& \multirow{3}*{$Y_{s2}^{J=0}$} &1.9/-1.66/2.12&\multirow{3}*{$Y_t^{J=0}$}&2.5/-2.23/1.82& \multirow{3}*{$Y_{d1}^{J=0}$}&3.3/-0.94/2.61&\multirow{3}*{$Y_{d2}^{J=0}$}&3.4/-1.62/2.03\\
                            &&1.30/-8.82/1.25&                    &1.95/-6.65/1.13&&2.7/-8.04/1.01&&3.4/-5.10/1.14&&3.5/-6.53/1.03\\
                            &&1.35/-19.47/0.94&                    &2.00/-17.77/0.74&&2.9/-18.64/0.69&&3.5/-12.41/0.75&&3.6/-14.64/0.70\\
\cline{2-11}

& \multirow{3}*{$Y_{s1}^{J=1}$} &1.25/-3.59/1.76&\multirow{3}*{$Y_{s2}^{J=1}$}&1.96/-1.05/2.64& \multirow{3}*{$Y_{t}^{J=1}$} &4.27/-3.20/1.53&\multirow{3}*{$Y_{d1}^{J=1}$}&4.98/-1.08/2.46\\
                  &                               &1.30/-8.98/1.22&&2.05/-6.92/1.13&                    &4.54/-10.85/0.87&&4.99/-1.27/2.26\\
                  &                              &1.35/-17.40/0.95&&2.14/-18.81/0.74&                   &4.81/-24.74/0.60&&5.00/-1.49/2.10\\
\cline{2-11}
& \multirow{3}*{$Y_{s1}^{J=2}$} &0.96/-1.28/2.59&\multirow{3}*{$Y_{s2}^{J=2}$}&2.0/-2.50/1.80\\
                  &                               &1.05/-7.99/1.25&&2.1/-7.25/1.14\\
                  &                                &1.14/-20.12/0.90&&2.2/-14.63/0.87\\
\bottomrule[1pt]
\end{tabular}
\end{center}
\end{table*}

\begin{center}
\begin{figure*}[htbp!]
\begin{tabular}{c}
\includegraphics[bb=10 175 580 550,scale=0.7,clip]{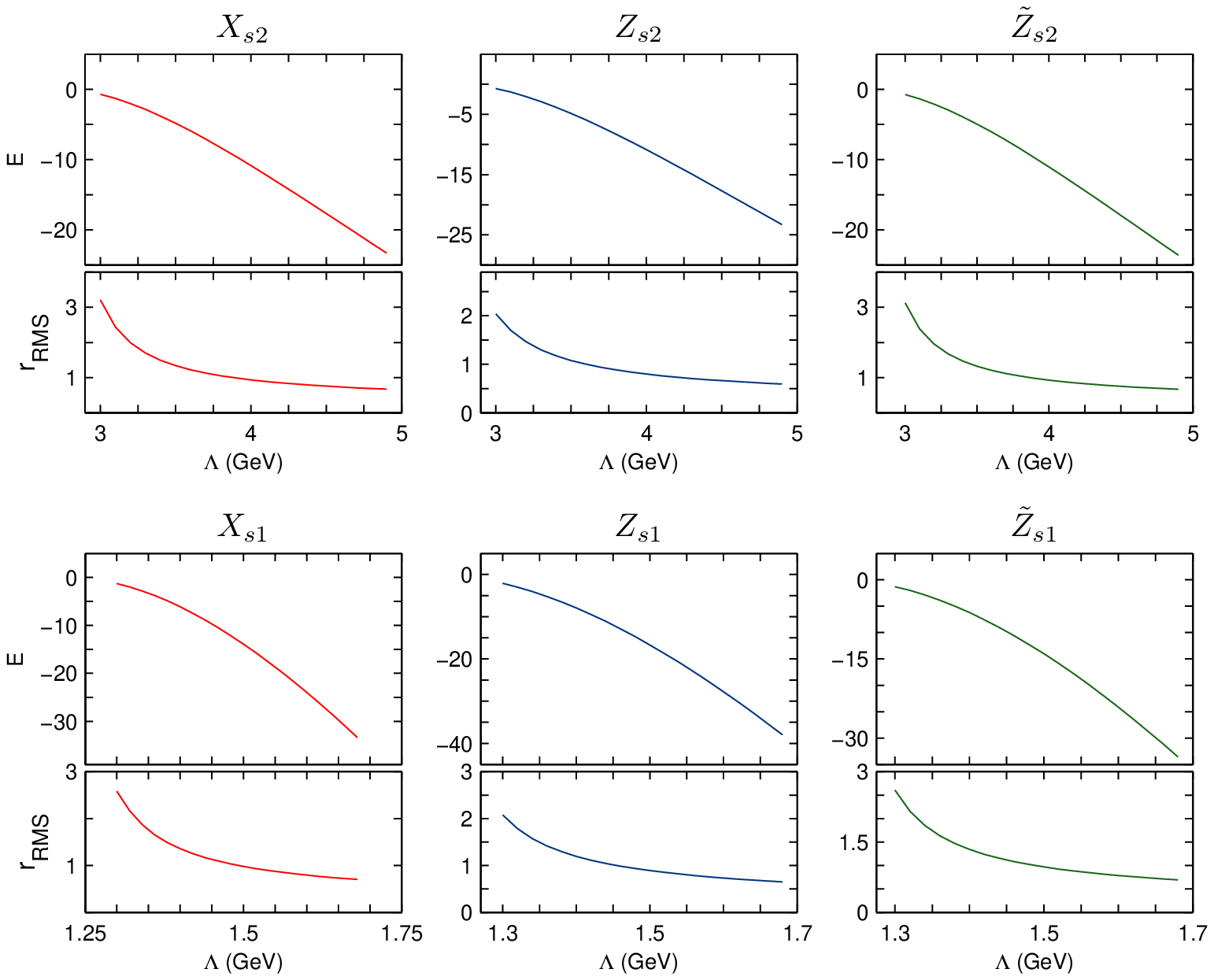}\\
\includegraphics[bb=20 170 580 650,scale=0.86,clip]{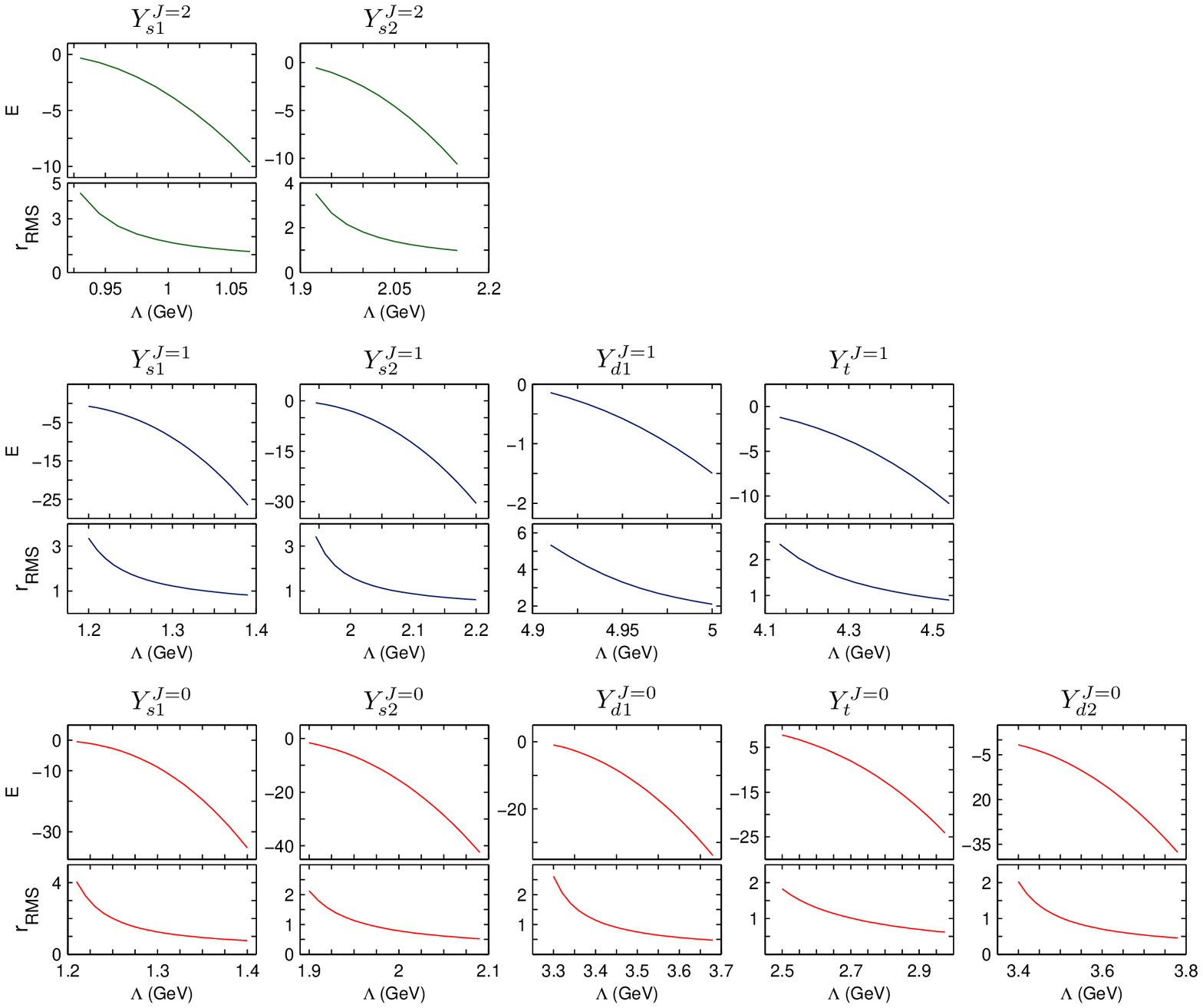}
\end{tabular}
\caption{(color online). The variation of the binding energy $E$
and root-mean-square radius $r_{RMS}$ with $\Lambda$ for the
$\mathcal{D}^{(*)}\mathcal{B}^{(*)}$ system. Here $E$ and
$r_{RMS}$ are in units of MeV and fm. The superscript $J=0$,
$J=1$, $J=2$ denotes the total angular momentum $J$. \label{p1}}
\end{figure*}
\end{center}

In Table \ref{tp}, we list the obtained typical values of the
bound-state solution of these $B_c$-like molecular systems, while
the dependence of the results on $\Lambda$ is given in Fig.
\ref{p1}. Among the 24 cases shown in Table. \ref{potential},
we find that there exist the bound-state solutions only for 17
states:

\begin{enumerate}

\item{$\mathcal{DB}$: We find the bound-state solution only for
the $X_{s1}$ and $X_{s2}$ states. Both of these states are of the
same quantum number, {\it i.e.}, $I(J^P)=0(0^+)$. The values of
the cutoff $\Lambda$ is close to 1 GeV for the $X_{s1}$ state. For
the other isosinglet $X_{s2}$, the bound-state solution appears
when taking $\Lambda\sim 3.2$ GeV.}

\item{$\mathcal{D^*B}/\mathcal{DB^*}$: The bound-state solution
exists only for the four isosinglets $Z_{s1}$, $Z_{s2}$,
$\tilde{Z}_{s1}$ and $\tilde{Z}_{s2}$ with $0(1^+)$. Since the
effective potentials of the $\mathcal{D^*B}$ and $\mathcal{DB^*}$
systems are the same, the dependence of the bound solutions on
$\Lambda$ for $Z_{s1}$ and $Z_{s2}$ are almost similar to those of
$\tilde{Z}_{s1}$ and $\tilde{Z}_{s1}$ respectively (see Fig.
\ref{p1}). The small difference of the reduced masses also results
in the difference of the typical values listed in Table. \ref{tp}
when comparing the results of the states marked by the same
subscript $s1$ or $s2$.
    }

\item{$\mathcal{D^*B^*}$: For the $\mathcal{D^*B^*}$ systems,
there are 15 states. Among them we find 11 states with bound-state
solutions, which include the isosinglets $Y_{s1}^{J=0}$,
$Y_{s2}^{J=0}$, $Y_{s1}^{J=1}$, $Y_{s2}^{J=1}$, $Y_{s1}^{J=2}$,
$Y_{s2}^{J=2}$, isodoublets $Y_{d1}^{J=0}$, $Y_{d2}^{J=0}$,
$Y_{d1}^{J=1}$, and isodoublets $Y_{t}^{J=0}$, $Y_{t}^{J=1}$. }

\end{enumerate}

We use a hand-waving notation, i.e., five-star, four-star,
three-star and two-star etc to mark the states in order to
indicate that the bound-state solutions exist when the cutoff
parameter $\Lambda$ corresponds to the different values:
$\Lambda<1.5$ GeV, $1.5<\Lambda<2.5$ GeV, $2.5<\Lambda<3.5$ GeV,
$3.5<\Lambda<5$ GeV respectively. In this way we categorize these
states according to the numerical results listed in Fig. \ref{p1}
and Table \ref{tp} (see Table \ref{remark} for more details).
Usually the cutoff $\Lambda$ is taken around 1 GeV, which is a
reasonable value, especially in the deuteron case. Thus, a
five-star state implies that a loosely molecular state probably
exists. {The mass spectra of the
$B\bar{D}$, $B\bar{D}^*$, $B^*\bar{D}$, and $B^*\bar{D}^*$
molecular states with the $\{Q\bar{q}\}\{\bar{Q}^{(\prime)}q\}$
configuration were studied with the QCD sum rule approach
\cite{Zhang:2009vs}, which correspond to the above six five-star
$B_c$-like molecular states obtained in this work. }

\begin{table*}[htb]
\begin{center}
%\vspace{-3mm}
%\footnotesize
\caption{Summary of the $B_c$-like systems. \label{remark}}
\begin{tabular}{ccclccclcclccccccc}
\toprule[1pt]
 &State&$I(J^P)$&Remark           & &State&$I(J^P)$&Remark         &State&$I(J^P)$&Remark\\
\midrule[1pt]
\multirow{2}*{$\mathcal{DB}$}& $X_{s1}$ & $0(0^+)$&$\star\star\star\star\star$&   & $Y^{J=0}_{s1}$ & $0(0^+)$&$\star\star\star\star\star$&   $Y^{J=1}_{s2}$ & $0(1^+)$&$\star\star\star\star$\\
& $X_{s2}$ & $0(0^+)$&$\star\star\star$&& $Y^{J=0}_{s2}$ & $0(0^+)$&$\star\star\star\star$&   {$Y^{J=1}_{s1}$} & $1(1^+)$&$\star\star$  \\
\multirow{2}*{$\mathcal{D^*B}$}& $Z_{s1}$ & $0(1^+)$&$\star\star\star\star\star$&\multirow{2}*{$\mathcal{D^*B^*}$} & $Y^{J=0}_{t}$ & $1(0^+)$&$\star\star\star\star$        &$Y^{J=1}_{d1}$ & $\frac{1}{2}(1^+)$&$\star\star$\\
& $Z_{s2}$ & $0(1^+)$&$\star\star\star$&        & $Y^{J=0}_{d1}$ & $\frac{1}{2}(0^+)$&$\star\star\star$&{$Y^{J=2}_{s1}$}&$0(2^+)$&$\star\star\star\star\star$ \\
\multirow{2}*{$\mathcal{DB^*}$}& $\tilde{Z}_{s1}$ & $0(1^+)$&$\star\star\star\star\star$&& $Y^{J=0}_{d2}$ & $\frac{1}{2}(0^+)$&$\star\star\star$&{$Y^{J=2}_{s2}$} &$0(2^+)$&$\star\star\star\star$\\
& $\tilde{Z}_{s2}$ & $0(1^+)$&$\star\star\star$& &{$Y^{J=1}_{s1}$} & $0(1^+)$&$\star\star\star\star\star$&&&&&\\
\bottomrule[1pt]
\end{tabular}
\end{center}
\end{table*}

In the following, we will discuss the allowed decay modes of these
predicted $B_c$-like molecular states that may be helpful to the
future experimental search. All the five-star states $X_{s1}$,
$Z_{s1}$, $\tilde{Z}_{s1}$, $Y_{s1}^{J=0}$, $Y_{s1}^{J=1}$ and
$Y_{s1}^{J=2}$ are the isosinglet with subscript $s1$. Their decay
modes are listed in the 2nd-7th columns of Table. \ref{decay}, respectively.
In addition, the decays of the four four-star states
$Y_{s2}^{J=0}$, $Y_{t}^{J=0}$, $Y_{s2}^{J=1}$ and $Y_{s2}^{J=2}$
are shown in the 8th-11th columns of Table. \ref{decay}, respectively. In Table. \ref{decay}, we also give
the decay modes of the remaining five three-star states. In these decay channels, the $B_c(1P_1)$ and
$B_c^\prime(1P_1)$ mesons are the mixture of the $1^1P_1$ and
$1^3P_1$ states \cite{Ebert:2011jc}: $|B_c(1P_1) \rangle =
|B_c(1^1P_1)\rangle \cos\theta + |B_c(1^3P_1)\rangle \sin\theta$,
$|B_c^\prime(1P_1) \rangle = -|B_c(1^1P_1)\rangle \sin\theta +
|B_c(1^3P_1)\rangle \cos\theta$. At present only $B_c(1^1S_0)$ was
observed with a mass $m(B_c(1^1S_0))=6277$ MeV
\cite{Nakamura:2010zzi}. We adopt the theoretical values from Ref.
\cite{Ebert:2011jc} when giving the decay channels of these
$B_c$-like molecular states, {\it i.e.}, $M_{B_c(1^3S_1)}=6333$
MeV, $M_{B_c(2^1S_0)}=6842$ MeV, $M_{B_c(2^3S_1)}=6882$ MeV,
$M_{B_c(1^3P_0)}=6699$ MeV, $M_{B_c(1P_1)}=6743$ MeV,
$M_{B_c^\prime(1P_1)}=6750$ MeV and $M_{B_c(1^3P_2)}=6761$ MeV
\cite{Ebert:2011jc}. In obtaining these decay channels, we have
only considered the ground state of the light meson.

\begin{table*}[htbp]
\begin{center}
\caption{The decay modes of the predicted $B_c$ like molecular states. Here, $\checkmark$ denotes that the corresponding decay mode is allowed. \label{decay}}
\begin{tabular}{c|cccccc|cccc|ccccccccccccccc}\toprule[1pt]
channels & $X_{s1}$ & $Z_{s1}$ & $\widetilde{Z}_{s1}$ & $Y^{J=0}_{s1}$ &$Y^{J=1}_{s1}$ &$Y^{J=2}_{s1}$& $Y_{s2}^{J=2}$ & $Y_t^{J=0}$ & $Y_{s2}^{J=1}$ & $Y_{s2}^{J=2}$ & $X_{s2}$ & $Z_{s2}$ & $\widetilde{Z}_{s2}$ & $Y^{J=0}_{d1}$ &$Y^{J=0}_{d2}$ \\\midrule[0.5pt]
$BD$&&&&\checkmark&&\checkmark&&\checkmark&&&&&&&\\
$BD^*$&&&&&\checkmark&\checkmark&&&&&&&&&\\
$B^*D$&&\checkmark&&&\checkmark&\checkmark&&&&&&&&&\\
$BD_s$&&&&&&&&&&&&&&\checkmark&\checkmark\\
$B_sD$&&&&&&&&&&&&&&\checkmark&\checkmark\\
$B_s^*D^*$&&&&&&&&&&&&&&\checkmark&\\
$B_sD_s$&&&&&&&\checkmark&&&\checkmark&&&&&\\
$B_sD_s^*$&&&&&&&&&\checkmark&\checkmark&&&&&\\
$B_s^*D_s$&&&&&&&&&\checkmark&\checkmark&&\checkmark&&&\\
$B_c(1^1S_0)\omega$&\checkmark&\checkmark&\checkmark&&\checkmark&\checkmark&&&&&&&&&\\
$B_c(1^3S_1)\omega$&\checkmark&\checkmark&\checkmark&\checkmark&\checkmark&\checkmark&&&&&&&&&\\
$B_c(1^1S_0)\eta^{(\prime)}$&&&&\checkmark&&\checkmark&\checkmark&&&\checkmark&\checkmark&&&&\\
$B_c(1^3S_1)\eta$&&\checkmark&\checkmark&&\checkmark&\checkmark&&&\checkmark&\checkmark&&\checkmark&\checkmark&&\\
$B_c(1^3S_1)\eta^{\prime}$&&&&&\checkmark&\checkmark&&&\checkmark&\checkmark&&\checkmark&\checkmark&&\\
$B_c(1^3P_0)\eta$&&\checkmark&&&\checkmark&&&&\checkmark&&&\checkmark&\checkmark&&\\
$B_c(1P_1)\eta$&&&&\checkmark&\checkmark&\checkmark&\checkmark&&\checkmark&\checkmark&\checkmark&\checkmark&\checkmark&&\\
$B_c^\prime(1P_1)\eta$&&&&\checkmark&\checkmark&\checkmark&\checkmark&&\checkmark&\checkmark&\checkmark&\checkmark&\checkmark&&\\
$B_c(1^3P_2)\eta$&&&&&\checkmark&\checkmark&&&\checkmark&\checkmark&&\checkmark&\checkmark&&\\
$B_c(2^1S_0)\eta$&&&&&&&\checkmark&&&\checkmark&&&&&\\
$B_c(2^3S_1)\eta$&&&&&&&&&\checkmark&\checkmark&&\checkmark&&&\\
$B_c(1^1S_0)\pi$&&&&&&&&\checkmark&&&&&&&\\
$B_c(2^1S_0)\pi$&&&&&&&&\checkmark&&&&&&&\\
$B_c(1P_1)\pi$&&&&&&&&\checkmark&&&&&&&\\
$B_c^\prime(1P_1)\pi$&&&&&&&&\checkmark&&&&&&&\\
$B_c(1^3S_1)\rho$&&&&&&&&\checkmark&&&&&&&\\
$B_c(1^1S_0)\phi$&&&&&&&&&\checkmark&\checkmark&&\checkmark&\checkmark&&\\
$B_c(1^3S_1)\phi$&&&&&&&\checkmark&&\checkmark&\checkmark&&\checkmark&\checkmark&&\\
$B_c(1^1S_0)\pi\pi$&&\checkmark&\checkmark&&\checkmark&&&&&&&&&&\\
$B_c(1^3S_1)\pi\pi$&\checkmark&\checkmark&\checkmark&\checkmark&\checkmark&\checkmark&&&&&&&&&\\
$B_c(1^3P_0)\pi\pi$&\checkmark&&&\checkmark&&\checkmark&&&&&&&&&\\
$B_c(1P_1)\pi\pi$&&\checkmark&\checkmark&&\checkmark&\checkmark&&&&&&&&&\\
$B_c^\prime(1P_1)\pi\pi$&&\checkmark&\checkmark&&\checkmark&\checkmark&&&&&&&&&\\
$B_c(1^3P_2)\pi\pi$&\checkmark&\checkmark&\checkmark&\checkmark&\checkmark&\checkmark&&&&&&&&&\\
$B_c(2^1S_0)\pi\pi$&&\checkmark&\checkmark&&\checkmark&&&&&&&&&&\\
$B_c(2^3S_1)\pi\pi$&&\checkmark&\checkmark&\checkmark&\checkmark&\checkmark&&&&&&&&&\\
$B_c(1^1S_0)K\bar{K}$&&\checkmark&&&\checkmark&&&&&&&&&&\\
$B_c(1^3S_1)K\bar{K}$&&&&\checkmark&\checkmark&\checkmark&&&&&&&&&
\\\bottomrule[1pt]
\end{tabular}
\end{center}
\end{table*}

\section{Discussion and conclusion}\label{sec4}

In short summary, we have studied the interaction between the
S-wave $D^{(*)}/D^{(*)}_s$ meson and S-wave $B^{(*)}/B^{(*)}_s$
meson in the OBE model. With the obtained effective potentials, we
predict the existence of many $B_c$-like molecular states where we
have already included the S-D mixing effect. Besides estimating
their mass spectrum, we also list their decay modes.

\begin{table}[htbp]
\begin{center}
%\vspace{-3mm}
%\footnotesize
\caption{The bound-state solutions of $Y_{s1}^{J=0}$, $Y_{s1}^{J=1}$ and $Y_{s1}^{J=2}$ only considering OPE potentials. Here, $E$, $\Lambda$, and $r_{\mathrm{RMS}}$ are in units of GeV, GeV, and fm, respectively.   \label{ope}}
\begin{tabular}{ccccccc}
\toprule[1pt]
State & $\Lambda/E/r_{\mathrm{RMS}}$ & State & $\Lambda/E/r_{\mathrm{RMS}}$ & State & $\Lambda/E/r_{\mathrm{RMS}}$ \\\midrule[0.5pt]
\multirow{3}*{$Y_{s1}^{J=0}$} &2.1/-2.45/2.06&\multirow{3}*{$Y_{s1}^{J=1}$}&2.4/-1.73/2.34&\multirow{3}*{$Y_{s1}^{J=2}$}&1.2/-2.96/1.77\\
&2.2/-6.25/1.40&&2.6/-6.55/1.35&&1.3/-7.78/1.18\\
&2.3/-12.61/1.06&&2.8/-15.81/0.96&&1.4/-15.53/0.89\\
\bottomrule[1pt]
\end{tabular}
\end{center}
\end{table}

For comparison, we list the bound-state solution for
$Y_{s1}^{J=0}$, $Y_{s1}^{J=1}$ and $Y_{s1}^{J=2}$ when considering
the one-pion-exchange (OPE) potential only in Table. \ref{ope}.
The one pion exchange force provides the main attraction in the
formation of the $B_c$-like molecular state, which is consistent
with the observation in Ref. \cite{Sun:2011uh}.

{For the other five-star states, we
find that there does not exist the bound-state solution only
considering the sigma meson exchange. Further, we notice that the
$\rho$ meson exchange plays a much more important role in the case of
the other five-star states. With $X_{s1}$ as an example, the
typical values of its bound-state solutions with the $\rho$ meson
exchange alone are $(\Lambda,E,r_{RMS})= (1.5,-1.52,2.43),
(1.6,-4.60,1.50), (1.7,-8.97,1.13)$. The comparison of these
results and those listed in Table \ref{tp} indeed indicates that
the $\rho$ meson exchange dominates the $X_{s1}$, where $E$,
$\Lambda$, and $r_{\mathrm{RMS}}$ are in units of GeV, GeV, and fm,
respectively.}

{With $Y_{s1}^{J=0}$ as an example,
we also examined the sensitivity of the results to the coupling
constant in the OPE case. When adopting $g=0.885$, which is 1.5
times larger than $g=0.59$ in Ref. \cite{Isola:2003fh}, we have to
lower the $\Lambda$ value in order to get the similar binding
energy to that in the case of taking $g=0.59$, {\it i.e.},
\begin{eqnarray}
 \left\{
  \begin{array}{ccc}
    E=-6.61\,\mathrm{MeV},&\Lambda=1.15 \,\mathrm{GeV},& g=0.885,\\
     E=-6.25\,\mathrm{MeV},&\Lambda=2.20 \,\mathrm{GeV},& g=0.56.\\
  \end{array}
\right.
\end{eqnarray}
Thus, the effect of varying the coupling constant on the bound-state solution can be compensated by changing the $\Lambda$
value.}

Most of the predicted $B_c$-like molecular states can decay into a
$B_c$ meson plus light mesons. It is possible to find these states
in the corresponding invariant mass spectrum. Recall that the
narrow resonance $X(3872)$ lies very close to the $D\bar{D}^*$
threshold, which was first observed in the $J/\psi \pi^+\pi^-$
invariant mass spectrum of the $B\to K J/\psi \pi^+\pi^-$ process
\cite{Choi:2003ue}. Similarly, the $Z_{s1}$ and $\tilde{Z}_{s1}$
states can decay into the $B_c(1^3S_1)\pi\pi$ mode. The $Y(3940)$
state was observed in $B\to K J/\psi \omega$ \cite{Abe:2004zs}
while $Y(4140)$ in $B\to K J/\psi \phi$ \cite{Aaltonen:2009tz}.
Similarly, the predicted $(Y_{s1}^{J=0}$, $Y_{s1}^{J=1}$,
$Y_{s1}^{J=2})$ or $(Y_{s2}^{J=0}$, $Y_{s2}^{J=1}$,
$Y_{s2}^{J=2})$ may be searched for in the $B_c(1^3S_1) \omega$ or
$B_c(1^3S_1)\phi$ modes, respectively.

In the future, it will also be important to calculate the branching
ratios of the different decay modes. Moreover, the investigation
of the $B_c$-like molecular states in other phenomenological
models is also very interesting. Hopefully the investigations
presented in this work will be useful to an experimental search of
them, which will be an interesting research topic.

\vfil

\section*{Acknowledgment} This
project is supported by the National Natural Science Foundation of
China under Grants 11175073, 11035006, 10625521,10721063, the
Ministry of Education of China (FANEDD under Grant No. 200924,
DPFIHE under Grant No. 20090211120029, NCET, the Fundamental
Research Funds for the Central Universities), the Fok Ying-Tong Education Foundation (No. 131006), the Ministry of
Science and Technology of China(2009CB825200) and CNPq and FAPESP -
Brazil.

\end{document}